\pdfoutput=1

\newcommand*{\ATLASLATEXPATH}{latex/}

\documentclass[cernpreprint,texlive=2016,txfonts,UKenglish]{\ATLASLATEXPATH atlasdoc}    

\usepackage[backend=bibtex]{\ATLASLATEXPATH atlaspackage}

\usepackage{\ATLASLATEXPATH atlasbiblatex}

\usepackage{\ATLASLATEXPATH atlasphysics}

\graphicspath{{logos/}{figures/}}

\usepackage{Moriond2017_HZZllvv_LowMass-defs}

\AtlasTitle{Search for an invisibly decaying Higgs boson or dark matter candidates produced in association with a 
$Z$ boson in $pp$ collisions at $\sqrt{s} =$ 13 \TeV~with the ATLAS detector}

\author{The ATLAS Collaboration}

\PreprintIdNumber{CERN-EP-2017-166}

\AtlasDate{\today}

\AtlasJournalRef{Phys. Lett. B 776 (2017) 318}

\AtlasDOI{10.1016/j.physletb.2017.11.049}

\AtlasAbstract{
  A search for an invisibly decaying Higgs boson or dark matter candidates produced in association with a leptonically decaying $Z$ boson in proton--proton collisions at $\sqrt{s} =$ 13 \TeV~is presented. This search uses 36.1 \ifb~of data collected by the ATLAS experiment at the Large Hadron Collider. No significant deviation from the expectation of the Standard Model backgrounds is observed. Assuming the Standard Model $ZH$ production cross-section, an observed (expected) upper limit of 67\% (39\%) at the 95\% confidence level is set on the branching ratio of invisible decays of the Higgs boson with mass $m_H = $ 125 \GeV. The corresponding limits on the production cross-section of the $ZH$ process with the invisible Higgs boson decays are also presented. Furthermore, exclusion limits on the dark matter candidate and mediator masses are reported in the framework of simplified dark matter models.
}

\hypersetup{pdftitle={ATLAS document},pdfauthor={The ATLAS Collaboration}}

\begin{document}

\maketitle

\section{Introduction}
\label{sec:intro}
The observation of the Higgs boson at the LHC~\cite{Aad:2012tfa,Chatrchyan:2012xdj} not only signified a success of the Standard Model (SM), but also opened a unique opportunity to search for new physics. In the SM, the invisible decay of the Higgs boson (\hzzinv) has a branching ratio  \invbr~of $1.06 \times 10^{-3}$ for $m_H = $ 125 \GeV~\cite{deFlorian:2016spz}. A larger \invbr~can exist in many extensions of the SM. For example, a Higgs boson can decay to light neutralinos~\cite{PhysRevD.37.719,Belanger:2001am}, graviscalars in extra-dimension models~\cite{Giudice:2000av,Battaglia:2004js}, Majorons~\cite{Shrock:1982kd,Joshipura:1992ua,Bonilla:2015uwa}, neutrinos~\cite{Shrock:1982kd,Belotsky:2002ym,Seto:2015rma}, or dark matter (DM) through the Higgs portal model~\cite{PhysRevD.62.041302,Englert:2011yb}. Observation of a \invbr~significantly above the SM value would give a strong indication for physics beyond the SM (BSM). 

The existence of DM is supported by a large body of astrophysical measurements, however its nature still remains mysterious. One of the hypotheses assumes that DM is composed of weakly interacting massive particles (WIMPs)~\cite{Lee:1977ua} that are nearly invisible to particle detectors. Experiments at the LHC can search for WIMPs produced in association with a detectable final state, and provide sensitive constraints on low-mass WIMP production~\cite{Malik:2014ggr,Buchmueller:2014yoa,Baudis:2016qwx}. Moreover, models with a sizable \invbr~often involve a Higgs boson decaying into WIMPs, and thus, studying \invbr~gives a unique probe into DM through its coupling to the Higgs boson.

The study of LEP data found no evidence of an invisibly decaying Higgs boson with $m_H < $ 114.4 \GeV~\cite{Searches:2001ab}, assuming a neutral CP-even Higgs boson produced at the SM rate and decaying with \invbr~$=$ 100\%. Both the ATLAS and CMS collaborations have extended the study to a higher mass range and reported their search results in multiple final states~\cite{Aad:2015txa,Aad:2015uga,Aad:2014iia,Aad:2015pla,Chatrchyan:2014tja,Khachatryan:2016whc,Khachatryan:2014jba,Khachatryan:2015vta}. Currently, the most stringent upper limit on \invbr~is around 24\% at the 95\% confidence level (CL)~\cite{Aad:2015pla,Khachatryan:2016whc} with $m_H = $ 125 \GeV. With certain assumptions, constraints on \invbr~can be inferred from the visible decay channels, and an upper limit of 34\% was obtained using LHC Run-1 data~\cite{Khachatryan:2016vau}. Similarly, DM has been searched for in a range of final states at the LHC~\cite{Aaboud:2017yqz,Aaboud:2016qgg,Aaboud:2017uak,Aad:2014vea,Aad:2014vka,Aaboud:2016tnv,Aaboud:2017dor,Khachatryan:2014tva,Sirunyan:2017hci,Khachatryan:2014rwa,Sirunyan:2017onm,Sirunyan:2017onm,Sirunyan:2017xgm,Sirunyan:2017hnk,Sirunyan:2017ewk,Khachatryan:2016mdm}, and no hints have been found to date. 

This Letter reports a search for an invisibly decaying Higgs boson with $m_H = $ 125 \GeV~or WIMPs produced in association with a $Z$ boson using 36.1 \ifb~of data collected by the ATLAS detector in 13 \TeV~$pp$ collisions. The search is carried out in a final state with two isolated electrons or muons from a $Z$ boson decay and large missing transverse momentum (\met) due to an invisible Higgs boson decay or a WIMP pair (\llmet). The BSM signal processes typically result in larger \met~than in background events. If no obvious deviation from the SM prediction is found, the observed \met~distribution is used to constrain the existence of new phenomena. An upper limit on \invbr~for $m_H = $ 125 \GeV~can be derived assuming the SM $ZH$ production cross-section. In simplified DM models~\cite{Buchmueller:2014yoa,Abercrombie:2015wmb,Abdallah:2015ter}, WIMP production is mediated by a spin-0 or spin-1 BSM particle (mediator) giving coupling constants to quarks (\gq) and WIMPs (\gchi). Fixing the coupling constants, exclusion limits on the WIMP mass (\mdm) and the mediator mass (\mmed) can be set. This search adopts a benchmark scenario where the WIMP pair is produced through the $s$-channel exchange of an axial-vector mediator. This choice is motivated by the findings in Ref.~\cite{Malik:2014ggr}, which indicated that LHC searches can be more sensitive than direct searches to WIMP production in this particular model with an axial-vector mediator. Figure~\ref{fig:diagram} gives the leading tree-level diagrams for both $ZH$ production and WIMP production in the benchmark model. 

\begin{figure}[!htbp]
\centering
\scalebox{0.7}{
    \includegraphics[width=0.49\linewidth]{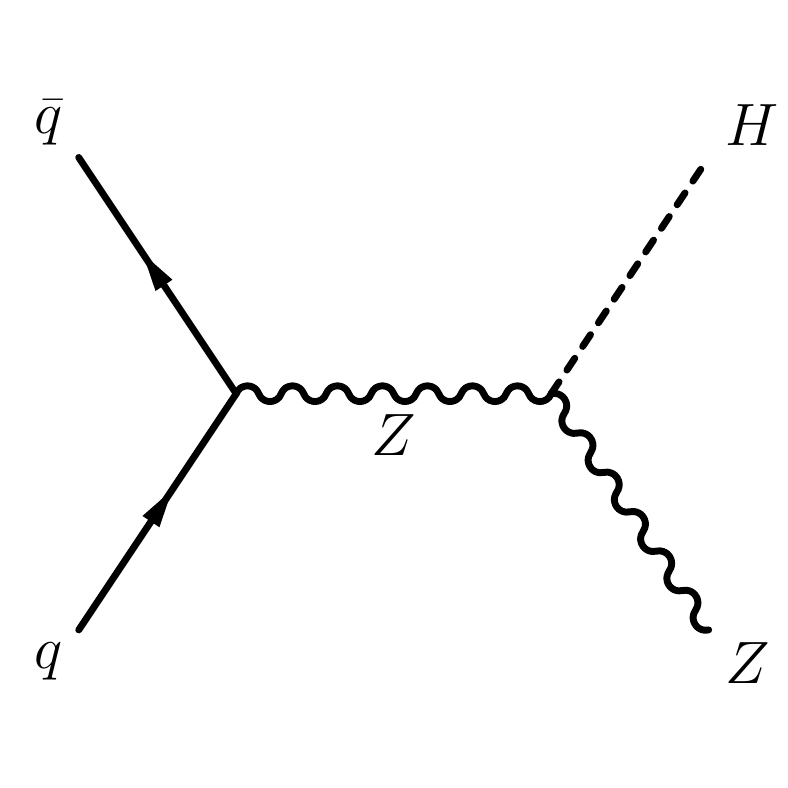}
    \includegraphics[width=0.49\linewidth]{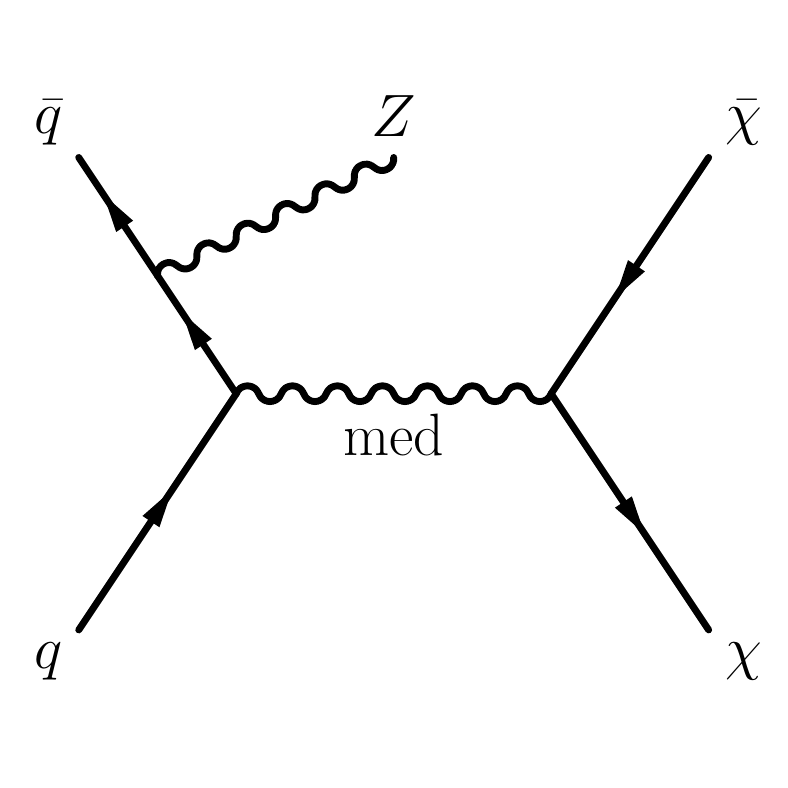}
    }
\caption{Leading tree-level diagrams for the $ZH$ production (left) and the WIMP pair production in the benchmark model (right).}
\label{fig:diagram}    
\end{figure}

\section{ATLAS detector}
\label{sec:detector}
The ATLAS detector~\cite{Aad:2008zzm,ATLAS-TDR-19} is a large multi-purpose apparatus with a forward-backward symmetric cylindrical geometry\footnotemark~and nearly 4$\pi$ coverage in solid angle. The collision point is encompassed by an inner tracking detector (ID) surrounded by a 2 T superconducting solenoid, electromagnetic (EM) and hadronic calorimeters, and a muon spectrometer (MS) with a toroidal magnetic field. The ID provides tracking for charged particles for $|\eta|<2.5$. It consists of silicon pixel and strip detectors surrounded by a straw tube tracker that also provides transition radiation measurements for electron identification. The EM and hadronic calorimeter system covers the pseudorapidity range $|\eta|<4.9$. For $|\eta|<2.5$, the liquid-argon EM calorimeter is finely segmented and plays an important role in electron and photon identification. The MS includes fast trigger chambers ($|\eta|<2.4$) and high-precision tracking chambers covering $|\eta|<2.7$.  A two-level trigger system selects events to be recorded for offline physics analysis~\cite{TRIG-2016-01}.

\footnotetext{ATLAS uses a right-handed coordinate system with its origin at the nominal interaction point (IP) in the centre of the detector and the $z$-axis along the beam pipe. The $x$-axis points from the IP to the centre of the LHC ring, and the $y$-axis points upward. Cylindrical coordinates $(r,\phi)$ are used in the transverse plane, $\phi$ being the azimuthal angle around the $z$-axis. The pseudorapidity is defined in terms of the polar angle $\theta$ as $\eta=-\ln\tan(\theta/2)$.} 

\section{Data and simulation}
\label{sec:datamc}
This search utilises data collected with single-lepton triggers by the ATLAS detector during the 2015 and 2016 data-taking periods. A combination of a lower \pt~threshold trigger with an isolation requirement and a higher \pt~threshold trigger without any isolation requirement is used. The \pt~threshold of the isolated electron (muon) trigger ranges from 24 (20) to 26 \GeV~depending on the instantaneous luminosity. The higher \pt~threshold is 50 (60) \GeV~for the electron (muon) case over all the data-taking periods. The overall trigger efficiency is above 98\% for the BSM signal processes after the full event selection described in Section~\ref{sec:selection}.

To study the invisible Higgs boson decays, Monte Carlo events are produced for the SM $ZH$ process with a subsequent $Z$ boson decay into a dilepton pair and the \hzzinv~decay (\zhsignal). The $ZH$ signal processes from both the quark--antiquark ($qqZH$) and gluon--gluon ($ggZH$) initial states are modelled with \progname{Powheg-Box} v2~\cite{powheg4,powheg5} using the CT10~\cite{ct10} parton distribution function (PDF) and interfaced to \progname{Pythia8.186}~\cite{Pythia8} for parton showering. The kinematic distributions of \zhsignal~events are described at next-to-leading-order (NLO) in QCD. Additionally, for the $qqZH$ process, the MINLO~\cite{MINLO} method is applied to improve the gluon resummation calculation, and the \ptz~distribution is corrected to NLO electroweak (EW) accuracy with a reweighting approach detailed in Ref.~\cite{deFlorian:2016spz}. The SM $ZH$ production cross-section is computed with next-to-next-to-leading-order (NNLO) QCD and NLO EW precision and found to be 884 fb~\cite{deFlorian:2016spz} with $m_H = $ 125 \GeV~at 13 \TeV. The DM signal is modelled with the leading-order \progname{MadGraph5\_aMC@NLO} matrix element~\cite{Alwall:2014hca} using NNPDF3.0~\cite{Ball:2014uwa} and showered with \progname{Pythia8.186}. 
DM signal events with an axial-vector mediator and fermionic WIMPs are produced for different \mmed~and \mdm, both in a range from 10 to 1000 \GeV. As recommended in Ref.~\cite{Abercrombie:2015wmb}, the DM events are generated by choosing \gq~$= 0.25$, \gchi~$= 1$, and a minimal mediator width. The AZNLO~\cite{Aad:2014xaa} and A14~\cite{ATL-PHYS-PUB-2014-021} parameter sets are used to tune the \progname{Pythia8.186} parton-shower for the simulation of the \zhsignal~and DM signals, respectively.

The backgrounds to this search include various diboson processes ($ZZ$, $WZ$, $WW$), the production of \ttbar, $Wt$, a $W$ or $Z$ boson in association with jets (\wjet, \zjet), and rare processes such as three-boson production (denoted by $VVV$ with $V = W$ or $Z$) and the production of \ttbar accompanied by one or two vector bosons ($t\bar{t}V(V)$). These background processes can result in the \llmet~final state with at least one boson decaying leptonically.

Production of $ZZ$ events is modelled with \progname{Powheg-Box} v2 and \progname{gg2vv3.1.6}~\cite{Kauer:2012hd,Kauer:2013qba} for the quark--antiquark ($qqZZ$) and gluon--gluon ($ggZZ$) initial states, respectively. The $qqZZ$ and $ggZZ$ events are described at NLO and LO QCD accuracies, respectively. The $qqZZ$ production cross-section is corrected to NNLO QCD and NLO EW precision using K-factors binned in the invariant mass of the $ZZ$ system, provided by the authors of Refs.~\cite{NNLOQCD,NLOEWcorr}. The QCD and EW corrections to the $qqZZ$ cross-section are assumed to factorise, as suggested in Ref.~\cite{Kallweit:2017khh}. In addition, the $ggZZ$ production cross-section is scaled to account for the NLO QCD correction~\cite{Caola:2015psa}. The $WZ$ and $WW$ processes are generated with \progname{Powheg-Box} v2, and their production cross-sections are predicted at NLO in QCD. All the diboson events are generated with the CT10 PDF set and showered using \progname{Pythia8.186} with the AZNLO tune. 

\progname{Sherpa2.2.1}~\cite{sherpa} is used to model the \zjet~process, and the $Z$ boson \pT distribution is matched to data. The \wjet~events are generated with \progname{Powheg-Box} v2 interfaced to \progname{Pythia8.186}. Both the \ttbar~and $Wt$ events are simulated with \progname{Powheg-Box} v2 and showered with \progname{Pythia6.428}~\cite{Sjostrand:2006za}. The cross-sections of these processes are all calculated at NNLO in QCD. The rare $VVV$ background, consisting of $WWW$, $WWZ$, $WZZ$ and $ZZZ$ production processes, is modelled with \progname{Sherpa2.1.1}. \progname{MadGraph5\_aMC@NLO} interfaced to \progname{Pythia8.186} is used to generate the $t\bar{t}V(V)$ background events that account for $\ttbar W$, $\ttbar Z$ and $\ttbar WW$ production processes. 
 
Generated events are processed through the ATLAS detector simulation~\cite{simuAtlas} based on GEANT4~\cite{Agostinelli:2002hh}. Additional $pp$ collisions in the same proton bunch crossing (pile-up) are simulated with \progname{Pythia8.186} and overlaid to simulated events to mimic the real collision environment. The distribution of the average number of interactions per bunch crossing in the simulation is weighted to reflect that in data. Simulated events are processed with the same reconstruction algorithms as for the data. Furthermore, the lepton momentum scale and resolution, the lepton reconstruction and identification efficiencies, and the trigger efficiencies in the simulation are corrected to match that measured in data.

\section{Selection criteria}
\label{sec:selection}
This search is carried out in a \llmet~final state, which contains large \met~and a pair of high-\pT~isolated electrons ($ee$) or muons ($\mu\mu$). 
Backgrounds are reduced by removing events with extra leptons or any jets containing $b$-hadrons (``$b$-jets'') and by requiring a boosted $Z$ boson which is back to back with the missing transverse momentum vector (\vecmet). Therefore, this search requires good measurement and identification of the leptons and jets and precise understanding of the \met. 

Electrons are reconstructed from energy deposits in the EM calorimeter matched to a track reconstructed in the ID. Candidate electrons must have \pT~$>$ 7 \GeV~and pseudorapidity $|\eta|<$ 2.47. Electrons must satisfy a set of likelihood-based identification criteria which are chosen to be approximately 90\% efficient and are referred to as the ``medium'' operating point~\cite{egamma}. Muons are reconstructed from a combined fit of tracks reconstructed independently in the ID and in the MS. 
Candidate muons must have \pT~$>$ 7 \GeV~and $|\eta|<$ 2.5. Muons are required to satisfy a set of identification criteria, 
which are referred to as the “medium” criteria~\cite{Aad:2016jkr}.
To suppress cosmic-ray and non-prompt contributions, the absolute value of the longitudinal impact parameter of leptons must be smaller than 0.5 mm, and the transverse impact parameter divided by its error must be less than 5 (3) for electrons (muons). ``Loose'' isolation criteria~\cite{Aad:2016jkr,egamma} are applied to remove jets misidentified as leptons or leptons from $b$-hadron decays, and the isolation selection varies as a function of \pT to maintain a uniform efficiency of 99\% for signal leptons. 

Jets are reconstructed with the anti-$k_t$ algorithm~\cite{antikt} with the radius parameter $R = 0.4$~\cite{jets,ATLAS-PHYS-PUB-2015-015,Cacciari:2011ma}. Candidate jets must have \pT~$>$ 20 \GeV~and $|\eta|<$ 4.5. Additional requirements using the track and vertex information inside a jet~\cite{jvt} are applied for jets with \pT~$<$ 60 \GeV~and $|\eta|<$ 2.5 to suppress pile-up contributions. Candidate $b$-jets ($\pt > 20$ \GeV~and $|\eta| < $ 2.5) are identified with an algorithm providing 85\% signal efficiency and a rejection factor of 33 for light-flavor jets~\cite{bjets}. The \vecmet~vector is computed as the negative of the vector sum of transverse momenta of all the leptons and jets, as well as the tracks originating from the primary vertex but not associated with any of the leptons or jets (``soft-term'')~\cite{Aad:2016nrq}. Usage of the track-based soft-term, rather than the calorimeter-based one, minimises the impact of pile-up on the \met~reconstruction.

Events are required to have a collision vertex associated with at least two tracks each with \pt~$>$ 0.4 \GeV. Candidate events must have exactly two selected electrons or muons with opposite charges and \pT~$>$ 20 \GeV, and the leading lepton is further required to have \pT~$>$ 30 \GeV. To suppress the $WZ$ background, events that contain an extra ``soft'' lepton are rejected, where the soft leptons satisfy the corresponding ``loose'' identification criteria and all other lepton selection criteria. The dilepton invariant mass ($m_{\ell\ell}$) is required to be in the range between 76 and 106 \GeV~to reject background processes with two leptons that do not originate from the prompt decay of a $Z$ boson (\nonresll). 

After the above selection (``preselection''), the data sample is still dominated by the \zjet~and \nonresll~processes, and further requirements on \met~and event topology are applied to suppress these backgrounds. Candidate events are required to have \met~$>$ 90 \GeV~and \metoht~$>$ 0.6, where \htdef~is calculated as the scalar sum of the \pT~of the selected leptons and jets. Since the signal processes tend to have a boosted $Z$ boson produced in the direction opposite to \vecmet, the azimuthal angle difference between the dilepton system and \vecmet, \dphizmet, must be larger than 2.7 radians, and the selected leptons must be close to each other, with \drll~$=\sqrt{(\dphill)^2 + (\detall)^2}<$ 1.8. Some of the remaining \zjet~background events have large \met~because of a significant soft-term contribution. To remove these \zjet~events, the absolute difference between the dilepton \pT~(\ptll) and the magnitude of the vector sum of \vecmet~and \ptvec~of all the selected jets (\ptmetjet) must be no more than 20\% of \ptll. Finally, events containing one or more $b$-jets are vetoed to suppress the \ttbar~and $Wt$ backgrounds. The event selection criteria are summarised in Table~\ref{tab:selection}. 

\begin{table}[!htbp]
\begin{center}
        \caption{Event selection criteria in the \llmet~search.}
        \label{tab:selection}
    \begin{tabular}{l c}
    \hline 
    \hline
                                                        &         Selection criteria \\
    \noalign{\smallskip}\hline \noalign{\smallskip}
    Two leptons                                & Two opposite-sign leptons, leading (subleading) $\pt > 30$ $(20)$ \GeV \\
    \noalign{\smallskip}\hline \noalign{\smallskip}
    Third lepton veto                       & Veto events if any additional lepton with $\pt > 7$ \GeV \\
    \noalign{\smallskip}\hline \noalign{\smallskip}
    $m_{\ell\ell}$                           & 76 $<m_{\ell\ell}<$ 106 \GeV \\
    \noalign{\smallskip}\hline \noalign{\smallskip}
    \met and \metoht                    & \met~$>$ 90 \GeV~and \metoht~$>$ 0.6 \\
    \noalign{\smallskip}\hline \noalign{\smallskip} 
     \dphizmet                                 &  \dphizmet~$>$ 2.7 radians \\
    \noalign{\smallskip}\hline \noalign{\smallskip}
     \drll                                 &  \drll~$<$ 1.8 \\    
     \noalign{\smallskip}\hline \noalign{\smallskip}
     Fractional \pt~difference & $\left | \ptll - \ptmetjet \right | / \ptll < $ 0.2 \\
     \noalign{\smallskip}\hline \noalign{\smallskip}
     $b$-jets veto & $N$($b$-jets) $= 0$ with $b$-jet \pt~$>$ 20 \GeV~and $|\eta|<$ 2.5 \\ 
     \noalign{\smallskip}\hline
    \hline
        \end{tabular}
\end{center}
\end{table}

The selection efficiency, defined as the product of the kinematic acceptance and the detector-level reconstruction and selection efficiency, is 10.0\% (10.6\%) for the \zhsignal~signal with $m_H = $ 125 \GeV~in the $ee$ ($\mu\mu$) channel. For a typical DM signal (\mmed~$=$ 500 \GeV~and \mdm~$=$ 100 \GeV) to which this search is sensitive, the efficiency is 13.4\% (13.7\%) for the $ee$ ($\mu\mu$) channel. The signal contribution from the $Z\rightarrow \tau\tau$ decay is found to be negligible, and therefore, only the prompt $Z\rightarrow ee$ ($Z \rightarrow \mu\mu$) decay is considered for the denominator in the efficiency calculation for the $ee$ ($\mu\mu$) channel.

\section{Uncertainties and background estimation}
\label{sec:background}
The selection efficiencies for the signal processes are subject to theoretical and experimental uncertainties. These systematic uncertainties are also evaluated for the \met~distributions, which are used to constrain the existence of new phenomena in this search.

The theoretical uncertainties originate from the PDF choice, the perturbative calculation, and the parton-shower modelling. These uncertainties are estimated in the same manner for both the \zhsignal~and DM signals. The PDF uncertainty covers the 68\% CL eigenvector uncertainty~\cite{ct10,Ball:2014uwa} of the nominal PDF set used in generating the signal events, as well as the difference between the nominal and alternative PDF sets. The alternative PDF sets used for the \zhsignal~(DM) signal are NNPDF3.0 and MSTW2008NLO~\cite{Martin:2009iq} (CT14lo~\cite{Dulat:2015mca} and MMHT2014lo68cl~\cite{Harland-Lang:2014zoa}). The perturbative uncertainty covers the variations from changing the QCD renormalisation and factorisation scales independently by factors ranging from one half to two. The parton-shower uncertainty is evaluated by varying parameters in the parton shower tunes according to Refs.~\cite{Aad:2014xaa, ATL-PHYS-PUB-2014-021}. In addition, the uncertainty in the NLO EW correction to the \ptz~distribution is considered for the \zhsignal~process. The total theoretical uncertainty is around 5\% on the selection efficiencies of both the \zhsignal~and DM signals. The SM $ZH$ production cross-section is assumed in the study of \invbr, and an uncertainty of 5\%~\cite{deFlorian:2016spz} is assigned to this prediction. The theoretical uncertainties on the signal \met~distributions are found to be minor.

The major experimental uncertainties relate to the luminosity uncertainty, the momentum scale and resolution of leptons and jets, and the lepton reconstruction and selection efficiencies. Smaller experimental uncertainties that are also considered include uncertainties due to the trigger selection efficiency, the determination of the \met~soft-term, the pile-up correction, and the $b$-jet identification efficiency. All the experimental uncertainties are included in the simulation-based predictions of the signal efficiencies, background yields, and \met~shapes. Overall, the total experimental uncertainty on the signal selection efficiency is around 5\%, dominated by the jet, lepton and pile-up components. The uncertainty on the combined 2015 and 2016 integrated luminosity is 3.2\%, derived following a methodology similar to that detailed in Ref.~\cite{Aaboud:2016hhf}, from a preliminary calibration of the luminosity scale using x--y beam-separation scans performed in August 2015 and May 2016. The luminosity uncertainty is considered for the background contributions estimated from simulation and for the \zhsignal~signal prediction when studying \invbr. 

Background contributions are either estimated from simulation or determined using data, as described below. Production of $ZZ$ events constitutes the dominant fraction (59\%) of the total background. Some $WZ$ events can be selected if the $W$ boson decay results in an electron or muon escaping detection or a hadronically decaying $\tau$, and this background accounts for 25\% of the total background. The \zjet~process with the $Z$ boson decaying to an $ee$ or $\mu\mu$ pair and poorly reconstructed \met~amounts to about 8\% of the total background, and a similar contribution originates from the \nonresll~processes consisting of \ttbar, $Wt$, $WW$ and $Z\rightarrow\tau\tau$ production. Minor contributions ($< 1 \%$) are expected from the \wjet, $VVV$, and $t\bar{t}V(V)$~backgrounds.

In this search, the $ZZ$ background is estimated from simulation, because the data sample with four charged leptons, which could be used to constrain the $ZZ$ background normalisation, is statistically limited. Overall, the NNLO QCD ($\approx+$10\%) and NLO EW corrections ($\approx-$10\%) to the $qqZZ$ yield are found to cancel each other out. The perturbative uncertainty and the PDF uncertainty (estimated as the CT10 eigenvector uncertainty at the 68\% CL) on the $qqZZ$ yield are estimated using the simulated sample, which has NLO accuracy in QCD. These uncertainties are found to be 4\% and 2\%, respectively. Both the perturbative and PDF uncertainties on the \met~shape are also considered for the $qqZZ$ process. In addition, a smaller uncertainty due to the parton-shower modelling is also assigned to the $qqZZ$ yield. An uncertainty of 60\% is assigned to the $ggZZ$ yield to cover the perturbative uncertainty on the NLO correction to the production cross-section and the theoretical uncertainty on the selection acceptance. The total experimental uncertainty on the $ZZ$ estimate is about 7\%, and the total uncertainty amounts to 10\%.

The $WZ$ background contribution predicted by simulation is scaled by a data-driven scale factor that accounts for potential missing higher-order calculations in the simulation. To derive the scale factor, a data control region enriched in $WZ$ events is defined with the preselection criteria, except that a third lepton with \pt$>$ 20 \GeV~and satisfying the medium identification criteria is allowed. In addition, a requirement of \mtW~$>$ 60 \GeV~is imposed in the control region to suppress non-$WZ$ contributions, where \mtW~is constructed from the third lepton's momentum and the \vecmet~vector. The scale factor is then calculated in the control region as the number of data events, after subtracting the non-$WZ$ contributions (estimated from simulation), divided by the predicted $WZ$ yield, and is found to be 1.29. The statistical uncertainty on the $WZ$ estimate is about 2\%, due to the limited size of the data control sample. The systematic uncertainty is evaluated for the ratio of the simulated $WZ$ yields in the signal and control regions. The experimental uncertainty on this ratio is about 4\%, while the theoretical uncertainty is negligible. The total uncertainty on the $WZ$ estimate is about 5\%. Moreover, theoretical uncertainties on the simulation-based \met~shape due to PDF and QCD scales are taken into consideration for the $WZ$ process.

A data-driven method is used to estimate the \zjet~background. This method defines three independent $Z$-enriched regions (B, C and D) that are disjoint from the signal region A. Then the data yields after subtracting the non-$Z$ contributions in these regions (\nB, \nC~and \nD) are used to predict the \zjet~contribution in the signal region (\nA), calculated as \BCD. An intrinsic assumption of \ABoCD~is made for the \zjet~process. To ensure that this assumption is valid, the control regions are defined so as to have the closure factor \ABCD~close to unity. The control regions are defined after the preselection, and a requirement of \met~$>$ 60 \GeV~and \metoht~$>$ 0.12 (``cleaning cut'') is imposed to remove the low-\met~phase space that is far away from the signal region. Since the \met~and the topological variables used in the event selection are expected to have only a small correlation, they are used to define regions B, C and D. Events are sorted into region B if \met~$<$ 90 \GeV~or \metoht~$<$ 0.6 and into region C if satisfying both the \met~and \metoht~selections but failing to satisfy any of the remaining criteria, and the rest of the events constitute region D. The closure factor \ABCD~is estimated using the simulated \zjet~events and found to be 1.3 (1.1) for the $ee$ ($\mu\mu$) final state, and both factors are consistent with unity, considering the large statistical uncertainties of the simulated samples and the experimental uncertainties. The major uncertainties on the \zjet~estimate include the difference between the closure factor and unity (``non-closure'') and the experimental and modelling uncertainties on the closure factor. The experimental uncertainty on the closure factor is dominated by the uncertainties on the jet energy scale and resolution. The modelling uncertainty covers the variations from changing the cleaning cut's values conservatively by 40\%. Smaller uncertainties due to the statistical uncertainty of the data and the subtraction of non-$Z$ contributions in the control regions are also considered. A total uncertainty of $^{+90\%}_{-55\%}$ ($^{+37\%}_{-49\%}$) is assigned to the \zjet~estimate in the $ee$ ($\mu\mu$) channel. Overall, the \zjet~background contribution in the $ee$ channel has a larger uncertainty than in the $\mu\mu$ channel, due to the larger non-closure and the larger modelling uncertainties in the $ee$ channel. Additionally, an alternative method, which corrects the simulated \zjet~contribution in the signal region by a data-driven scale factor derived in a sideband region defined by reversing the \metoht~cut, yields a consistent result. The \met~distribution for the \zjet~background is derived from simulation, and the shape uncertainty includes the experimental uncertainties and the difference between the simulated \met~distribution and that observed in data with \metoht~$<$ 0.6. 

To estimate the \nonresll~background, a control region dominated by the \nonresll~processes is defined by applying all the event selection criteria to the final state with an opposite-sign $e\mu$ pair and large \met. The \nonresll~contribution in the $ee$ ($\mu\mu$) channel is calculated as one half of the observed data yield after subtracting the contribution from the other background processes in the control region, and then corrected for the difference in the lepton reconstruction and identification efficiencies between selecting an $e\mu$ pair and an $ee$ ($\mu\mu$) pair. The lepton efficiency correction is derived as the square root of the ratio of the numbers of $\mu\mu$ and $ee$ events in data after the preselection, and this correction is obtained as a function of \pt~and $\eta$ of both leptons.
The total uncertainty on the \nonresll~estimates is about 14\%, including the statistical uncertainty of the data in the control region (13\%) and the method bias estimated from simulation (5\%). The \met~distributions for the \nonresll~background are derived from the data control region, and the differences between data and simulation are taken as the shape uncertainty.

The $VVV$ and $t\bar{t}V(V)$ backgrounds are estimated from simulation, and their contributions have a total uncertainty of about 20\%, including both the theoretical cross-section~\cite{Maltoni:2015ena,Binoth:2008kt} and experimental uncertainties. The \wjet~background is estimated using the fake-factor method described in Ref.~\cite{Aad:2011kk}.

\section{Result and interpretations}
\label{sec:result}
Table~\ref{tab:result} gives the observed data yields, the estimated background contributions, and the expectations for the two signal processes after the final selection. The observed and predicted \met~distributions in the $ee$ and $\mu\mu$ channels are shown in Figure~\ref{fig:final_met}. No significant excess over the SM background expectation is observed. 

\begin{table}[!htbp]
\begin{center}
        \caption{Observed data yields and expectations for the signal and background contributions in the signal region.
        The first error is statistical, and the second systematic. 
         The \zhsignal~signal contribution is shown with \invbr~$=0.30$, which is the value most compatible with data. 
         The DM signal contribution with \mmed~$=$ 500 \GeV~and \mdm~$=$ 100 \GeV~is 
         also scaled (with a factor of 0.27) to the best-fit contribution.
        The background contributions from the \wjet~, $VVV$ and $t\bar{t}V(V)$ processes are summed and presented with the label ``Others''. 
        The systematic uncertainty on the \zjet~contribution is taken as its upper systematic error.
        The uncertainty on the total background prediction is quadratically summed from those on the individual background contributions.}
        \label{tab:result}
    \begin{tabular}{l c c}
    \hline
    \hline
        Final State & $ee$ & $\mu\mu$ \\
    \hline
        Observed Data & 437 & 497 \\
    \hline
        Signal & & \\
        \zhsignal~(\invbr~$=$ 30\%) & 32 $\pm$ 1 $\pm$ 3  & 34 $\pm$ 1 $\pm$ 3 \\
        DM (\mmed~$=$ 500 \GeV, \mdm~$=$ 100 \GeV) $\times$ 0.27
                          & 10.8 $\pm$ 0.3 $\pm$ 0.8  & 11.1 $\pm$ 0.3 $\pm$ 0.8 \\                          
    \hline
        Backgrounds & & \\
        $qqZZ$    & 212 $\pm$ 3 $\pm$ 15  & 221 $\pm$ 3 $\pm$ 17 \\
        $ggZZ$    & 18.9 $\pm$ 0.3 $\pm$ 11.2  & 19.3 $\pm$ 0.3 $\pm$ 11.4 \\
        $WZ$  & 106 $\pm$ 2 $\pm$ 6  & 113 $\pm$ 3 $\pm$ 5 \\
        \zjet     & 30 $\pm$ 1 $\pm$ 28  & 37 $\pm$ 1 $\pm$ 19 \\
        \Nonresll & 30 $\pm$ 4 $\pm$ 2  & 33 $\pm$ 4 $\pm$ 2 \\
        Others & 1.4 $\pm$ 0.1 $\pm$ 0.2  & 2.5 $\pm$ 2.0 $\pm$ 0.8 \\
    \hline
        Total Background & 399 $\pm$ 6 $\pm$ 34 & 426 $\pm$ 6 $\pm$ 28 \\
    \hline
    \hline
        \end{tabular}
\end{center}
\end{table}

\begin{figure}[!htbp]
\centering
    \includegraphics[width=0.49\linewidth]{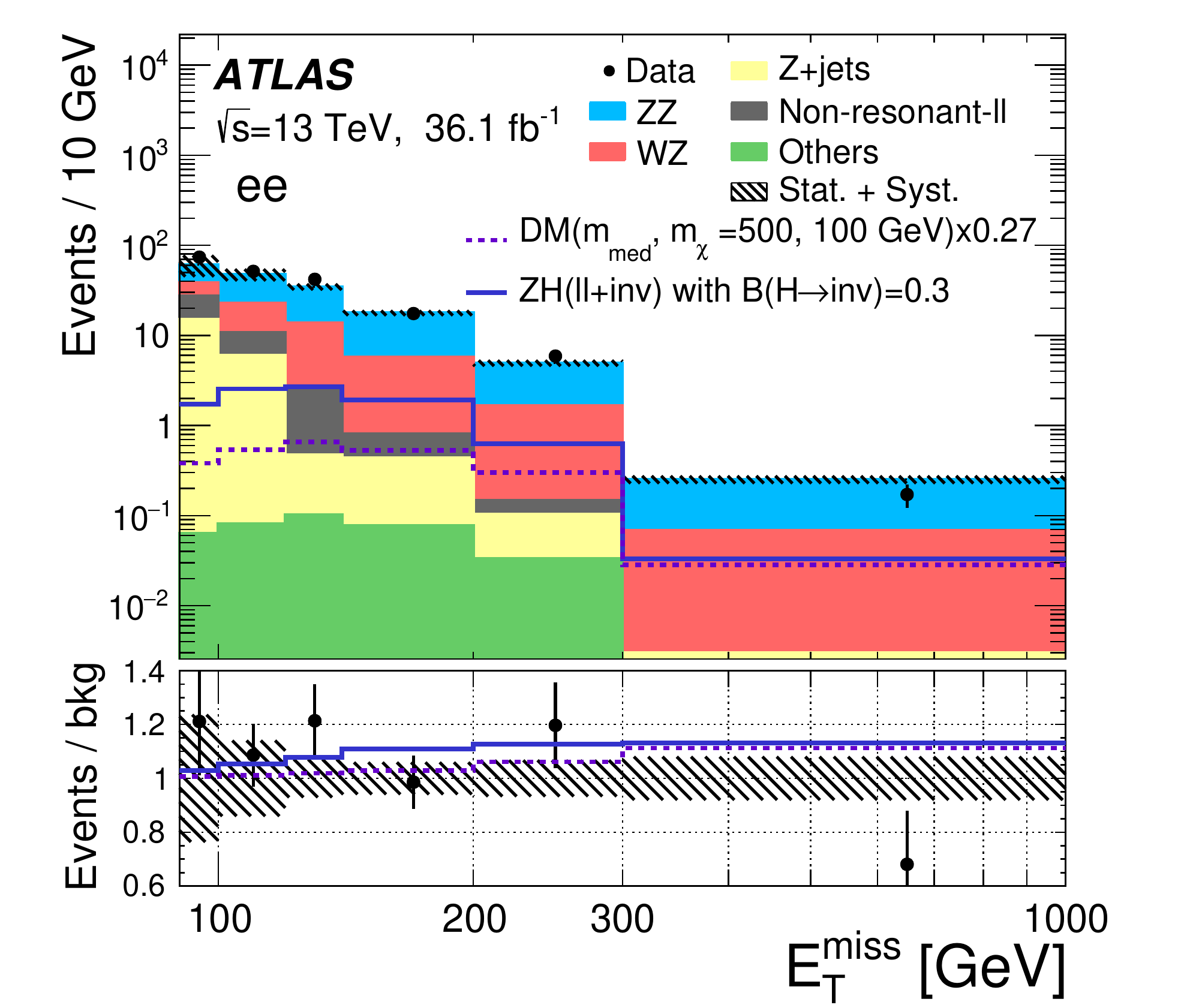}
    \includegraphics[width=0.49\linewidth]{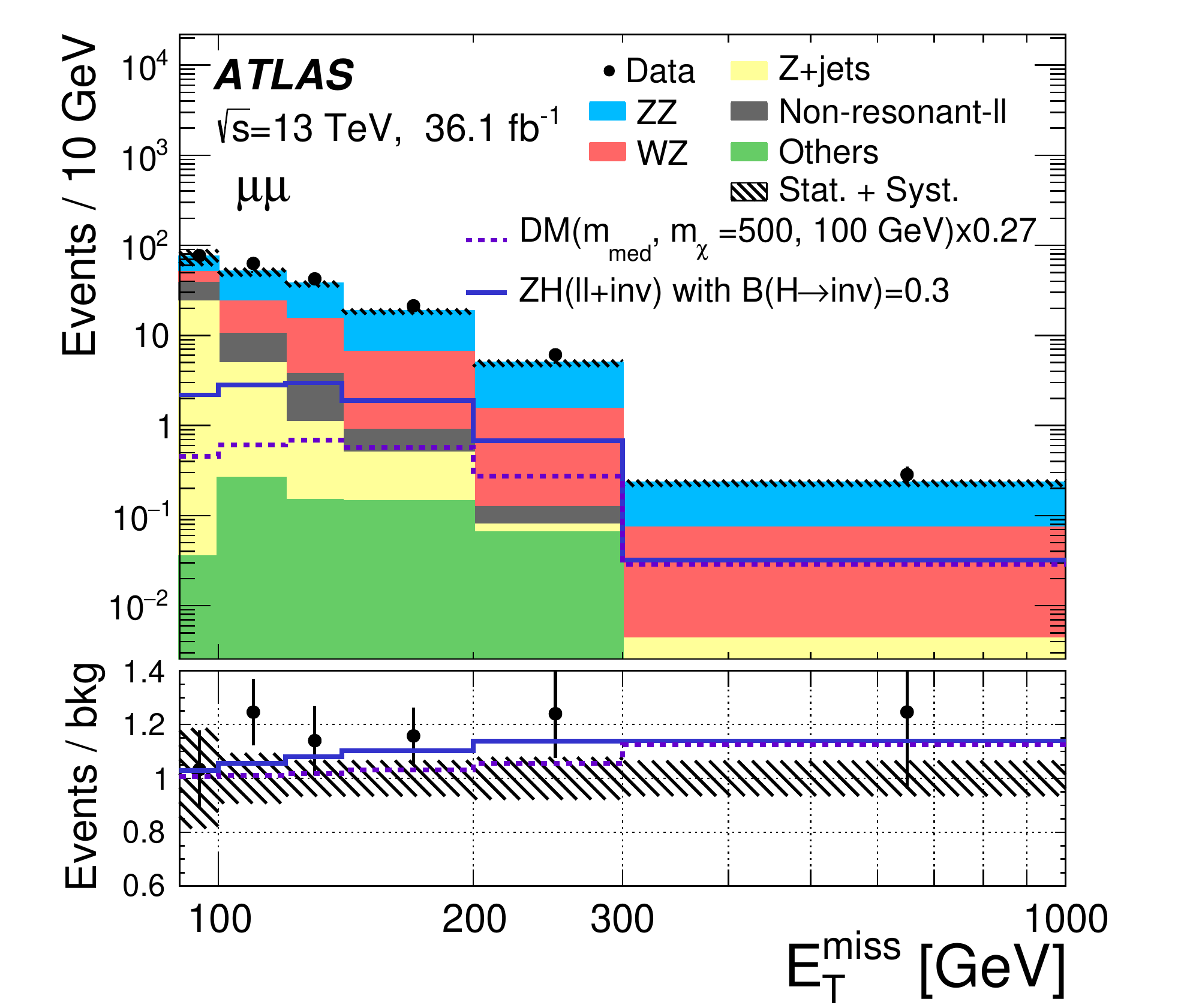}
\caption{Observed \met~distribution in the $ee$ (left) and  $\mu\mu$ (right) channel compared to the signal and background predictions.
    The error band shows the total statistical and systematic uncertainty on the background prediction.
    The background predictions are presented as they are before being fit to the data.
    The ratio plot gives the observed data yield over the background prediction (black points) as well as 
    the signal-plus-background contribution divided by the background prediction (blue or purple line) 
    in each \met~bin.
    The rightmost bin contains the overflow contributions.
    The \zhsignal~signal distribution is shown with \invbr~$=0.3$, which is the value most compatible with data. 
    The simulated DM distribution with \mmed~$=$ 500 \GeV~and \mdm~$=$ 100 \GeV~is 
    also scaled (with a factor of 0.27) to the best-fit contribution.
    }
\label{fig:final_met}    
\end{figure}

To examine the compatibility of the data and the signal-plus-background hypothesis, a test statistic is defined using the profile likelihood ratio method~\cite{Cowan:2010js}. The likelihood function is the product of all the Poisson probability density functions built in individual \met~bins and final states. In each bin the observed number of events in data is represented by a Poisson probability density function with a mean equal to the sum of the predicted signal and background yields. The systematic uncertainties are implemented as nuisance parameters (NPs) constrained by auxiliary Gaussian functions. In most cases, a common NP is used to account for each systematic uncertainty in all the \met~bins and in both the $ee$ and $\mu\mu$ channels. The statistical uncertainty on the \zjet~estimate is treated as being uncorrelated between the $ee$ and $\mu\mu$ channels, and the statistical uncertainties of the simulated samples are uncorrelated among all bins and final states. A frequentist method with the CLs formalism~\cite{cls} is then applied to set upper limits on the overall signal contribution, which is the parameter of interest left free in the test statistic.

There is a small data excess in the $\mu\mu$ channel, and the $p$-value for the compatibility of the data and the background-only hypothesis is 0.014, which corresponds to a significance of about $2.2\sigma$. Combining the $ee$ and $\mu\mu$ channels, the $p$-value becomes 0.06 ($1.5\sigma$). Assuming the signal-plus-background hypothesis, the compatibility between the $ee$ and $\mu\mu$ channels is found to be $1.4\sigma$. 

Table~\ref{tab:invbr} gives the 95\% CL upper limits on \invbr, assuming the SM prediction for the $ZH$ production cross-section. As a result of the small data excess observed in this search, the observed limit is less stringent than the expected one. Using the combined $ee$ and $\mu\mu$ channel, the observed and expected limits on \invbr~are 67\% and 39\%, respectively. The corresponding observed (expected) limit on the production cross-section of the \zhsignal~process is 40 (23) fb at the 95\% CL, where only the prompt $Z\rightarrow ee$ and $Z\rightarrow \mu\mu$ decays are considered. When the signal-plus-background model is fit to the data, the best-fit \invbr~is $(30 \pm 20)\%$, where the data statistical and systematic uncertainties are about 13\% and 16\%, respectively. The dominant sources of the systematic uncertainty are the theoretical uncertainties on the $qqZZ$ and $ggZZ$ predictions, the luminosity uncertainty, the uncertainties in the data-driven estimation of the $WZ$ and \zjet~backgrounds, and the jet energy scale and resolution uncertainties.

\begin{table}[!htbp]
\begin{center}
        \caption{The 95\% CL upper limits on \invbr~for $m_H=$ 125 \GeV~from the $ee$, $\mu\mu$, and combined $ee+\mu\mu$ channels. 
        Both the observed and expected limits are given, and the $1\sigma$ and $2\sigma$ uncertainties 
        on the expected limits are also presented.}
        \label{tab:invbr}
    \begin{tabular}{l c c }
    \hline 
    \hline
                                &   Obs. \invbr~Limit &   Exp. \invbr~Limit $\pm 1\sigma \pm 2\sigma$   \\
     \hline \noalign{\smallskip}
    $ee$                   & 59\% & (51~$^{+21}_{-15}$~$^{+49}_{-24}$) \%  \\ 
    \noalign{\smallskip} \hline \noalign{\smallskip}
    $\mu\mu$        & 97\% & (48~$^{+20}_{-14}$~$^{+46}_{-22}$) \%  \\
    \noalign{\smallskip} \hline \noalign{\smallskip}
    $ee+\mu\mu$ & 67\%  & (39~$^{+17}_{-11}$~$^{+38}_{-18}$) \% \\    
    \noalign{\smallskip} \hline
    \hline
        \end{tabular}
\end{center}
\end{table}

Figure~\ref{fig:axialDM} gives the 95\% CL exclusion limit in the two-dimensional phase space of WIMP mass \mdm~and mediator mass \mmed~derived using the combined $ee+\mu\mu$ channel, where the underlying dark matter model assumes an axial-vector mediator, fermionic WIMPs, and a specific scenario of the coupling parameters (\gq~$= 0.25$, \gchi~$= 1$). From the observed limits at the 95\% CL, the mediator mass \mmed~is excluded up to 560 \GeV~for a light WIMP, while the WIMP mass \mdm~is excluded up to 130 \GeV~for \mmed~$=$ 400 \GeV. For the bulk of the phase space, the observed limit is weaker than the expected one by about 1$\sigma$. The compatibility of the observed and expected limits is better than that for the \invbr~limits, mainly because the sensitivity region for the DM signals has larger \met~and the difference between the observed yield and the background expectation is less statistically significant at high \met. 

\begin{figure}[!htbp]
\centering
    \includegraphics[width=0.49\linewidth]{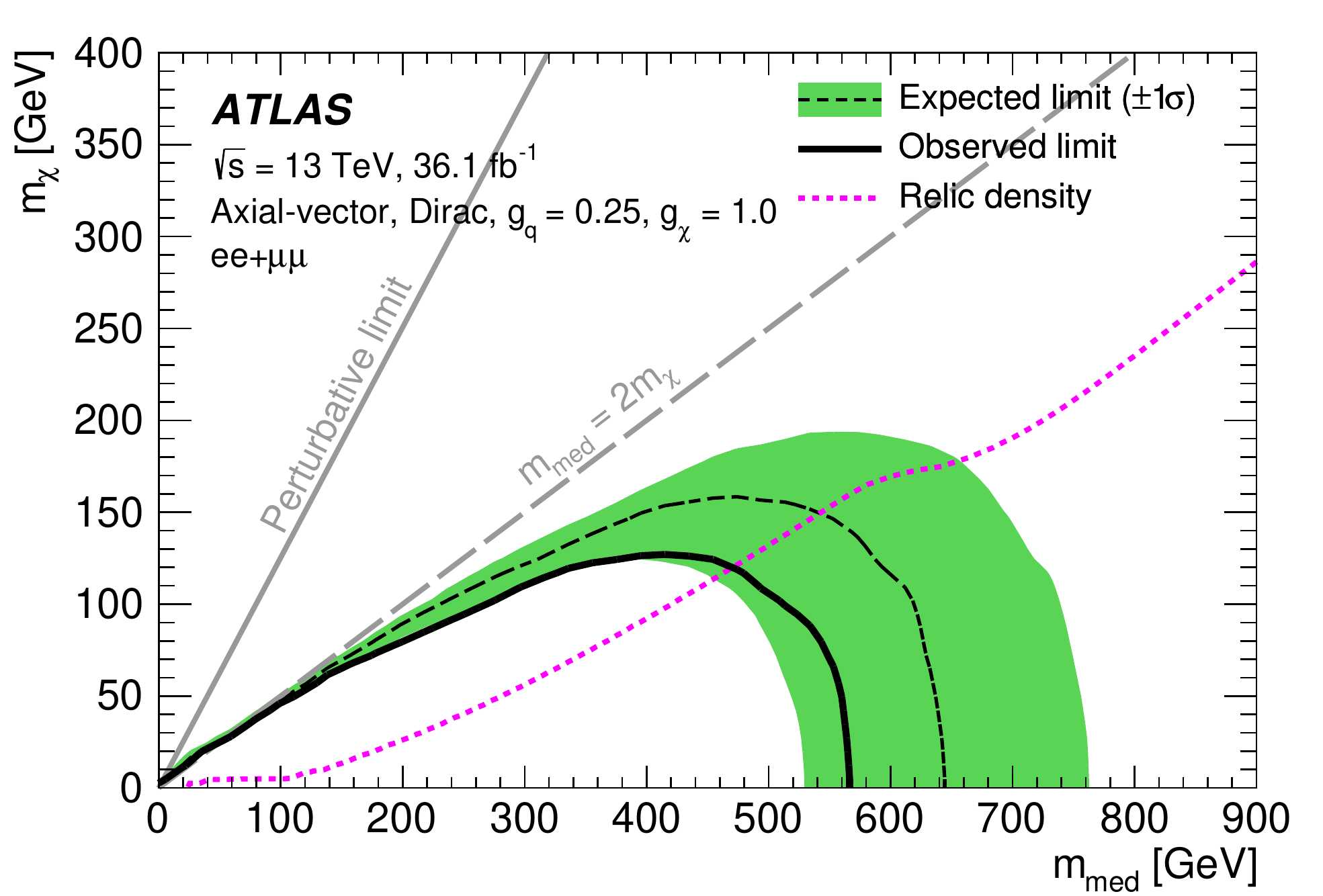}
\caption{DM exclusion limit in the two-dimensional phase space of WIMP mass \mdm~vs mediator mass \mmed~determined using the combined $ee+\mu\mu$ channel. Both the observed and expected limits are presented, and the $1\sigma$ uncertainty band for the expected limits is also provided. Regions bounded by the limit curves are excluded at the 95\% CL. The grey line labelled with ``\mmed~$=$~2\mdm'' indicates the kinematic threshold where the mediator can decay on-shell into WIMPs, and the other grey line gives the perturbative limit~\cite{Boveia:2016mrp}. The relic density line~\cite{Boveia:2016mrp} illustrates the combination of \mdm~and \mmed~that would explain the observed DM relic density.
}
\label{fig:axialDM}    
\end{figure}

\FloatBarrier

\section{Conclusion}
\label{sec:conclusion}
This Letter presents a search for an invisibly decaying Higgs boson or WIMPs produced in association with a $Z$ boson using 36.1 \ifb~of data collected by the ATLAS detector in $pp$ collisions at $\sqrt{s} = $ 13 \TeV~at the LHC. The search is carried out in the \llmet~final state. There is no significant data excess above the expectation of the SM backgrounds. An observed (expected) upper limit of 67\% (39\%) is set on \invbr~at the 95\% CL for $m_H = $ 125 \GeV, which can be compared to the observed (expected) 95\% CL limit of 75\% (62\%) derived in the same final state using the ATLAS data collected at $\sqrt{s} = $~7 and 8 \TeV. The expected \invbr~limit is much improved compared to the previous one, while the improvement in the observed limit is marginal due to the small data excess observed in this search. The corresponding observed (expected) limit on the production cross-section of the $ZH$ process with prompt $Z\rightarrow ee$ and $Z\rightarrow \mu\mu$ decays and invisible Higgs boson decays is 40 (23) fb at the 95\% CL. Finally, exclusion limits are placed on masses in a simplified dark matter model with an axial-vector mediator and fermionic WIMPs. The mediator mass \mmed~is excluded up to 560 \GeV~at the 95\% CL for a light WIMP, while the WIMP mass \mdm~is excluded up to 130 \GeV~ for \mmed~$=$ 400 \GeV. The constraint on the existence of dark matter from this search provides another input to the global search for dark matter at the LHC.

\section*{Acknowledgements}

We thank CERN for the very successful operation of the LHC, as well as the
support staff from our institutions without whom ATLAS could not be
operated efficiently.

We acknowledge the support of ANPCyT, Argentina; YerPhI, Armenia; ARC, Australia; BMWFW and FWF, Austria; ANAS, Azerbaijan; SSTC, Belarus; CNPq and FAPESP, Brazil; NSERC, NRC and CFI, Canada; CERN; CONICYT, Chile; CAS, MOST and NSFC, China; COLCIENCIAS, Colombia; MSMT CR, MPO CR and VSC CR, Czech Republic; DNRF and DNSRC, Denmark; IN2P3-CNRS, CEA-DSM/IRFU, France; SRNSF, Georgia; BMBF, HGF, and MPG, Germany; GSRT, Greece; RGC, Hong Kong SAR, China; ISF, I-CORE and Benoziyo Center, Israel; INFN, Italy; MEXT and JSPS, Japan; CNRST, Morocco; NWO, Netherlands; RCN, Norway; MNiSW and NCN, Poland; FCT, Portugal; MNE/IFA, Romania; MES of Russia and NRC KI, Russian Federation; JINR; MESTD, Serbia; MSSR, Slovakia; ARRS and MIZ\v{S}, Slovenia; DST/NRF, South Africa; MINECO, Spain; SRC and Wallenberg Foundation, Sweden; SERI, SNSF and Cantons of Bern and Geneva, Switzerland; MOST, Taiwan; TAEK, Turkey; STFC, United Kingdom; DOE and NSF, United States of America. In addition, individual groups and members have received support from BCKDF, the Canada Council, CANARIE, CRC, Compute Canada, FQRNT, and the Ontario Innovation Trust, Canada; EPLANET, ERC, ERDF, FP7, Horizon 2020 and Marie Sk{\l}odowska-Curie Actions, European Union; Investissements d'Avenir Labex and Idex, ANR, R{\'e}gion Auvergne and Fondation Partager le Savoir, France; DFG and AvH Foundation, Germany; Herakleitos, Thales and Aristeia programmes co-financed by EU-ESF and the Greek NSRF; BSF, GIF and Minerva, Israel; BRF, Norway; CERCA Programme Generalitat de Catalunya, Generalitat Valenciana, Spain; the Royal Society and Leverhulme Trust, United Kingdom.

The crucial computing support from all WLCG partners is acknowledged gratefully, in particular from CERN, the ATLAS Tier-1 facilities at TRIUMF (Canada), NDGF (Denmark, Norway, Sweden), CC-IN2P3 (France), KIT/GridKA (Germany), INFN-CNAF (Italy), NL-T1 (Netherlands), PIC (Spain), ASGC (Taiwan), RAL (UK) and BNL (USA), the Tier-2 facilities worldwide and large non-WLCG resource providers. Major contributors of computing resources are listed in Ref.~\cite{ATL-GEN-PUB-2016-002}.

\printbibliography

\newpage 
\begin{flushleft}
{\Large The ATLAS Collaboration}

\bigskip

M.~Aaboud$^\textrm{\scriptsize 137d}$,
G.~Aad$^\textrm{\scriptsize 88}$,
B.~Abbott$^\textrm{\scriptsize 115}$,
O.~Abdinov$^\textrm{\scriptsize 12}$$^{,*}$,
B.~Abeloos$^\textrm{\scriptsize 119}$,
S.H.~Abidi$^\textrm{\scriptsize 161}$,
O.S.~AbouZeid$^\textrm{\scriptsize 139}$,
N.L.~Abraham$^\textrm{\scriptsize 151}$,
H.~Abramowicz$^\textrm{\scriptsize 155}$,
H.~Abreu$^\textrm{\scriptsize 154}$
% ...
\bigskip
\\
% ...
$^{12}$ Institute of Physics, Azerbaijan Academy of Sciences, Baku, Azerbaijan\\
% ...
$^{88}$ CPPM, Aix-Marseille Universit{\'e} and CNRS/IN2P3, Marseille, France\\
% ...
$^{115}$ Homer L. Dodge Department of Physics and Astronomy, University of Oklahoma, Norman OK, United States of America\\
% ...
$^{119}$ LAL, Univ. Paris-Sud, CNRS/IN2P3, Universit{\'e} Paris-Saclay, Orsay, France\\
% ...
$^{137}$ $^{(a)}$ Facult{\'e} des Sciences Ain Chock, R{\'e}seau Universitaire de Physique des Hautes Energies - Universit{\'e} Hassan II, Casablanca; $^{(b)}$ Centre National de l'Energie des Sciences Techniques Nucleaires, Rabat; $^{(c)}$ Facult{\'e} des Sciences Semlalia, Universit{\'e} Cadi Ayyad, LPHEA-Marrakech; $^{(d)}$ Facult{\'e} des Sciences, Universit{\'e} Mohamed Premier and LPTPM, Oujda; $^{(e)}$ Facult{\'e} des sciences, Universit{\'e} Mohammed V, Rabat, Morocco\\
% ...
$^{139}$ Santa Cruz Institute for Particle Physics, University of California Santa Cruz, Santa Cruz CA, United States of America\\
% ...
$^{151}$ Department of Physics and Astronomy, University of Sussex, Brighton, United Kingdom\\
% ...
$^{154}$ Department of Physics, Technion: Israel Institute of Technology, Haifa, Israel\\
$^{155}$ Raymond and Beverly Sackler School of Physics and Astronomy, Tel Aviv University, Tel Aviv, Israel\\
% ...
$^{161}$ Department of Physics, University of Toronto, Toronto ON, Canada\\
% ...
$^{a}$ Also at Department of Physics, King's College London, London, United Kingdom\\
$^{b}$ Also at Institute of Physics, Azerbaijan Academy of Sciences, Baku, Azerbaijan\\
$^{c}$ Also at Novosibirsk State University, Novosibirsk, Russia\\
$^{d}$ Also at TRIUMF, Vancouver BC, Canada\\
$^{e}$ Also at Department of Physics {\&} Astronomy, University of Louisville, Louisville, KY, United States of America\\
% ...
$^{*}$ Deceased
\end{flushleft}

\end{document}